\newcommand{\templateusepackage}[1]{\usepackage{template/#1}}
\documentclass{template/egpubl}
\ConferencePaper

\usepackage[T1]{fontenc}
\usepackage{template/dfadobe}  

\usepackage{cite}  %

\BibtexOrBiblatex
\electronicVersion
\PrintedOrElectronic
\ifpdf \usepackage[pdftex]{graphicx} \pdfcompresslevel=9
\else \usepackage[dvips]{graphicx} \fi

\templateusepackage{egweblnk} 

\usepackage{booktabs} %

\usepackage{color}
\usepackage{bm}
\usepackage{soul}
\usepackage{enumitem}
\usepackage{hyperref}
\usepackage{gensymb}
\usepackage{amsmath}
\usepackage{makecell}

\usepackage[ruled]{algorithm2e} %

\SetAlFnt{\small}
\SetAlCapFnt{\small}
\SetAlCapNameFnt{\small}
\SetAlCapHSkip{0pt}

\newif\ifsubmit
\submitfalse

\newcommand{\beq}{\begin{equation}}
\newcommand{\eeq}{\end{equation}}

\newcommand{\hpart}{part}

\newcommand{\hparts}{\hpart{}s}

\newcommand{\rig}{rig}

\newcommand{\rigging}{rigging}

\title[How to Train Your Dragon]%
      {How to Train Your Dragon: Automatic Diffusion-Based Rigging for Characters with Diverse Topologies}

\author[Z. Gu et al.]
{\parbox{\textwidth}{\centering Zeqi Gu$^{1,3}$\orcid{0009-0002-5777-4560}
        Difan Liu$^{2}$\orcid{0000-0001-5971-2748} 
       Timothy Langlois$^2$\orcid{0000-0002-5043-8698}
       Matthew Fisher$^2$\orcid{0000-0002-8908-3417}
       Abe Davis$^3$\orcid{0000-0003-1469-2696}
        }
        \\
{\parbox{\textwidth}{\centering $^1$Cornell Tech, USA\\
         $^2$Adobe Research, USA\\
       $^3$ Cornell University, USA
       }
}
}

\begin{document}

\teaser{
  \includegraphics[width=\textwidth]{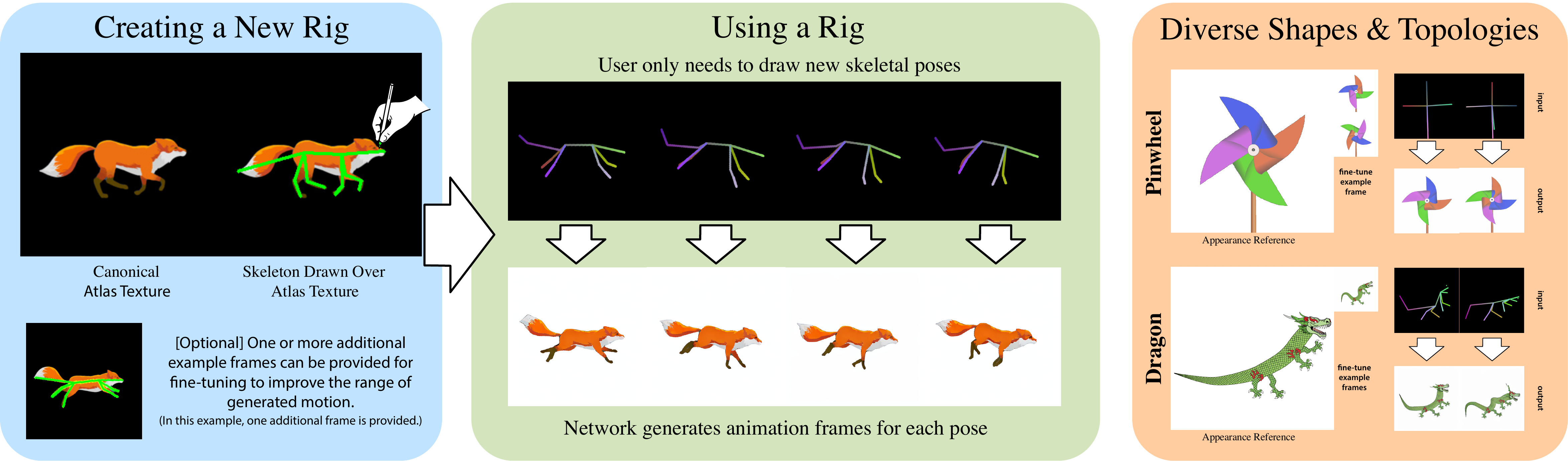}
 \centering
  \caption{\textbf{Workflow.} We propose AniDiffusion, a diffusion-based automatic rigging method that helps cartoon creators re-pose a character without going through the traditional arduous rigging procedure. Our method first provides an annotation user interface and asks the user to provide several keyframes of the character with desired control points annotated. Then we invite the user to input the coordinates of those keypoints to indicate how they want to re-pose the character per-frame. After a simple fine-tuning, our model is able to synthesize the character in the new motion. Unlike most existing works, our method does not assume the topology nor the texture of the character belonging to a certain category such as humanoids, and thus works on a much wider range of objects, such as a fox (blue and green box), a pinwheel (orange box, top row), and a dragon (orange box, bottom row). Image credits: fox \copyright Edina Gecse and pinwheel \copyright Ekaterine Kantaria; the dragon is self-created.}
\label{fig:teaser}
}

\maketitle
\begin{abstract}
   Recent diffusion-based methods have achieved impressive results on animating images of human subjects. However, most of that success has built on human-specific body pose representations and extensive training with labeled real videos. In this work, we extend the ability of such models to animate images of characters with more diverse skeletal topologies.
Given a small number (3--5) of example frames showing the character in different poses with corresponding skeletal information, our model quickly infers a rig for that character that can generate images corresponding to new skeleton poses.
We propose a procedural data generation pipeline that efficiently samples training data with diverse topologies on the fly. We use it, along with a novel skeleton representation, to train our model on articulated shapes spanning a large space of textures and topologies. Then during fine-tuning, our model rapidly adapts to unseen target characters and generalizes well to rendering new poses, both for realistic and more stylized cartoon appearances. 
To better evaluate performance on this novel and challenging task, we create the first 2D video dataset that contains both humanoid and non-humanoid subjects with per-frame keypoint annotations. With extensive experiments, we demonstrate the superior quality of our results.


\end{abstract}

\section{Introduction}
\label{sec:intro}
Animation has always been a labor-intensive process. Before the introduction of digital tools, animation was done by drawing each individual frame by hand, which meant an enormous amount of redundant work went into creating long sequences. Modern software helps reduce this redundant work by representing characters with a hierarchy of \hparts{} and corresponding transformations, which are collectively called a \emph{\rig{}}. Each \hpart{} in the \rig{} represents geometry that can appear in multiple frames, and each transformation represents an interpolatable way that the character can move. These rigs let artists animate a character by specifying the configuration of a corresponding skeletal representation at a sparse set of keyframes. This representation can greatly accelerate the animation workflow. However, the process of creating an animation \rig{}, called \emph{\rigging{}}, can be complicated and tedious. It involves factoring the character's geometry into \hparts{} and specifying how each \hpart{} is affected by transformations in the skeletal hierarchy, including the relative rigidity and bending of different parts in response to motion. To further complicate the process, much of a rig's information typically resides in metadata that the artist must define using specialized interfaces (e.g., layer decompositions and labels). In this work, we show how image diffusion networks can be trained to infer rigging behavior from a small set of example images and their corresponding skeletal structures. Based on this, we present a tool that greatly simplifies the 2D \rigging{} process. To create a 2D \rig{}, the artist only needs to trace the skeletal structure of one or more example frames showing the character. Then, the artist can draw new unseen skeletal poses, and our tool will generate corresponding images that are consistent with the degrees of freedom and style of deformations observed in the provided examples.

Ours is not the first work to create video by conditioning the generation of frames on a reference image and target poses. However, most work in this space has been limited to subjects with standard humanoid topologies--typically, the topology of human pose detectors (e.g., OpenPose~\cite{8765346} and DensePose~\cite{guler2018densepose}), which can be used to automatically label abundant training data. A distinct goal of our work is to build a solution that generalizes to more arbitrary and abstract characters, including humanoid characters with more diverse shapes, as well as animals and articulated objects with unseen topologies. In order to support such diversity, we need to adapt a new training strategy and pose representation that is less reliant on automatically-labeled video of humans.

Our work makes the following contributions:

\begin{itemize}
    \item We adapt a new pose conditioning strategy that generalizes to more diverse character appearance and skeletal topologies. Here we draw inspiration from the use of texture atlases and texture coordinates. 
    \item We present a training strategy that learns to generalize by training on procedurally generated synthetic data, and show that this strategy leads to generalization across real and cartoon images with little fine-tuning (approx. 25 minutes on an NVIDIA A100 GPU). 
    \item To foster research and better evaluation in this challenging task, we build the first 2D video dataset spanning both humanoids and non-humanoids with per-frame keypoint annotations to evaluate performance in this generalized problem setting.
    \item We demonstrate that our approach is able to effectively infer the parameters of classical non-neural rigging methods such as As-Rigid-As-Possible (ARAP)~\cite{sorkine2007rigid}, and Bounded Biharmonic Weights (BBW)~\cite{jacobson2011bounded} from the provided examples. This lets us understand our approach as a way to infer and use such rigging behavior from more general image inputs, including hand-drawn example frames.
\end{itemize}

\section{Related Works}
\subsection{Diffusion-Based Video Synthesis}
Diffusion models~\cite{song2020denoising,song2020score} were first developed to generate images~\cite{ramesh2022hierarchical,nichol2021glide}. As these image-based models demonstrate increasingly impressive power and are more computationally efficient than their video-based counterparts~\cite{ho2022video,ho2022imagen,saharia2022photorealistic}, how to adapt them for video generation has become an active research topic.
AnimateDiff~\cite{guo2023animatediff} expands the dimension on which the model operates from two to three by adding trainable weights for temporal attention between frames. After pre-training, this plug-in module can then be used with most image models to render consecutive dynamic frames. Our model is also image-based, and we include AnimateDiff as an optional element, though we find even without it our model is able to generate outputs stable across time.
A popular sub-topic of video generation is video editing, and the most common way to control the editing is via text prompts.
Tune-A-Video~\cite{wu2023tune} allows changing video content while preserving motions by finetuning a text-to-image diffusion model with a single text-video pair. FateZero~\cite{qi2023fatezero} proposes a training-free method by injecting the cross-attention map of the source video and modifying attention layers. Customize-a-Video~\cite{ren2024customize} and Lamp~\cite{wu2023lamp} learn the motion directly with example video(s).
These approaches could be viewed as transferring the motion from an original appearance to a new appearance, and differ from pose-conditioned method like ours in two crucial ways. First, the target appearance images are translated from the source image and the target motion is ideally the same as the input motion. Our method does not involve any image translation and the target motion is assumed to be different from the one in the input video. Second, such methods usually condition on a text prompt describing the goal of scene translation, and have no explicit modeling for the pose and geometry. Our work condition on the target appearance and a skeleton image instead of text, thus controlling the appearance and pose more directly. 

\subsubsection{Pose-Conditioned Animation}
Harnessing the generation power of diffusion models has been an active research direction. Many formats of conditioning have been proposed: Animate Anything~\cite{dai2023animateanything} uses a mask to indicate which part of the image needs to be animated. The motion is described by a text prompt, and the ``strength'' could roughly control the intensity of the motion. Other methods~\cite{yin2023dragnuwa, wu2025draganything, mou2023dragondiffusion, shi2024dragdiffusion} condition on a mouse drag that indicates the movement trajectory. Although these methods could animate objects other than humanoids, their controls are less precise as a trade-off for convenience. On the other hand, methods that enable more accurate pose controls have been focusing on humans, partially due to the lack of annotated data in other domains. These works usually acquire the skeleton annotations with an off-the-shelf human pose estimation method, such as DensePose~\cite{guler2018densepose} and OpenPose~\cite{cao2017realtime,wei2016cpm,simon2017hand,8765346}. 
DreamPose~\cite{karras2023dreampose} proposes an adapter to fuse the CLIP text embedding with the CLIP image embedding of the appearance image, and feed a projected version of the fused feature into the diffusion model for cross-attention. As their method was only trained on fashion datasets, the variation of appearances and poses is very limited. Later on, more generalizable approaches~\cite{xu2023magicanimate, zhong2024posecrafter,ma2024follow} emerged and could bu used zero-shot.
DisCo~\cite{wang2023disco} focuses on disentangled control of the foreground, background and pose, which enables human video generation with changeable foreground, background and motion. 
Animate Anyone~\cite{hu2023animate} designs a ReferenceNet to extract detail appearance features from reference images to serve as extra cross-attention values for the denoising UNet. The skeleton image is combined with the noisy latents to be the input. MagicPose~\cite{chang2023magicdance} is another state-of-the-art that shares similar design ideas and produces better facial expressions. We find the architecture of Animate Anyone achieves a balance between fine-tuning efficiency and performance, and thus adopt it for our method. As will be articulated in the following sections, our major differences from Animate Anyone are: (1) our skeleton representation is generalizable to diverse topologies and encodes depth ordering. (2) Our method is trained entirely on synthetic videos (before character-specific fine-tuning, which is required also for all compared methods once test characters are not confined to humanoids).

\subsection{Classical Rigging and Skinning}
Speeding up and improving the quality of rigging and skinning has been a long-standing topic in computer graphics. Some of the most fundamental algorithms include linear blend skinning, dual quaternions, and rigid skinning. After placing handles during the \textit{bind time}, most methods that are fast at \textit{pose time} compute the transformation at each object point by using a weighted blend of handle transformations. The specific optimization and weights vary across methods. Among methods that work on 2D and have code available, As-Rigid-As-Possible (ARAP)~\cite{sorkine2007rigid} and Bounded Biharmonic Weights (BBW)~\cite{jacobson2011bounded} are exemplary methods that have been widely adopted into animation engines. In Sec.~\ref{sec:results}, we will show that our model could infer a plausible interpolation between provided poses regardless of the underlying rendering engine that was used to generate the fine-tuning examples. Therefore, the user no longer needs to figure out the entire rig through trial and error -- they only need to annotate the keypoints on few frames, and let the diffusion model to implicitly do the reverse inference. This is one of our key strengths over non-neural methods, and with our ordering-aware skeleton bone representation, the user further avoids the need to manually separate the image frame into parts. Our model is able to infer the layer ordering and generate correct occlusions.

\section{Method}
\label{sec:method}
Our method involves two stages: (1) Stage 1 involves training on a large synthetic dataset to learn rigid rigging of diverse shapes, 
and (2) Stage 2 is fine-tuning on the given unseen test character (in Sec~\ref{sec:results}, we will show that even only trained on rigid deformations, our model can adapt to arbitrary non-rigid deformations after this quick fine-tuning).
We will first introduce our model architecture in Sec.~\ref{sec:preliminaries}. Then we describe the procedural generation of stage 1 training data: the appearance synthesis is in Sec.~\ref{sec:traindata_gen}, and pose representation is in Sec.~\ref{sec:skeleton_rep}. Training implementations of the two stages will be in the next section.

\begin{figure*}
\begin{center}
    \includegraphics[width=0.9\linewidth]{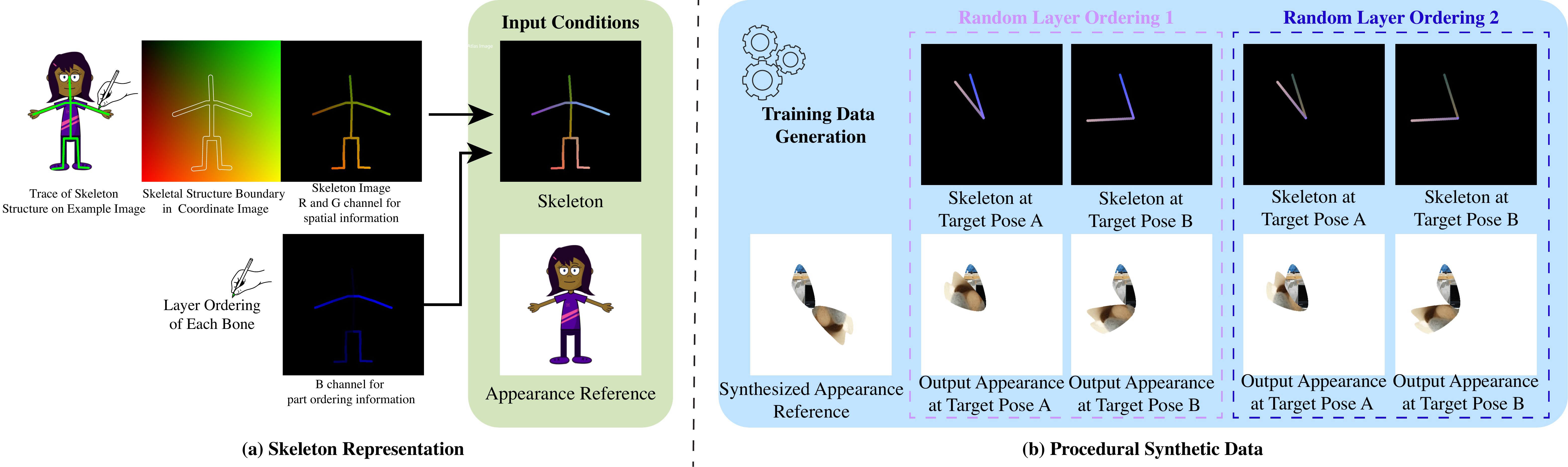}
\end{center}
    \caption{\textbf{Training Pipeline.} (a) Our model takes in an appearance reference image and a skeleton image as inputs. (b) For the first training stage, these are randomly generated through our data pipeline. With almost infinite possible combinations of texture, shape, and topology, our synthetic dataset is more challenging than any real-life datasets, which forces our model to learn the correct binding and deformations. Our skeleton representation for this wide range of topologies is also unique: in the Red and Green channel of this RGB image, we color pixels according to their $x$ and $y$ coordinates. When a user specifies a new target pose, this skeleton is transformed accordingly, which means that the value of pixels in the target skeleton image now refer to source coordinates in the starting rest pose. We use the Blue channel to embed layer ordering of each part of the body, which is crucial for characters that contain parts of different depths. For each appearance we train the model multiple target poses and layer orderings, as shown in the two dashed boxes in (b). When the new pose causes occlusions as in the two left columns, the supervising ground truth appearance is different when the order changes. Thus, our model is forced to understand the influence of layer ordering to appearance. For more data examples please refer to Fig.~\ref{fig:more_train}.}
\label{fig:networkinputs}
\end{figure*}

\subsection{Preliminaries on Diffusion Models}
\label{sec:preliminaries}
Diffusion models~\shortcite{ho2020denoising} have demonstrated impressive image generation capabilities through an iterative denoising process. Diffusion models consist of two Markov chains: a forward chain that perturbs data to noise, and a reverse chain that converts noise back to data.
For an input image $x_0$, the Gaussian noise is gradually added to $x_0$ through the forward Markov Chain:
\begin{equation}
    q(\mathbf{x}_t|\mathbf{x}_{t-1}) = \mathcal{N}(\mathbf{x}_t;\sqrt{1-\beta_t}\mathbf{x}_{t-1},\beta_t \mathbf{I})
\end{equation}
where $t = 1, ..., T$ denotes the timesteps, $\beta_t$ is a predefined variance schedule. At inference time, Gaussian noise is sampled from $\mathcal{N} (\mathbf{0}, \mathbf{I})$ and gradually denoised into the data distribution:
\begin{equation}
    p_\theta(\mathbf{x}_{t-1}|\mathbf{x}_t) = \mathcal{N}(\mathbf{x}_{t-1}; \boldsymbol{\mu}_\theta(\mathbf{x}_t, t), \sigma_t^2 \mathbf{I})
\end{equation}
where $\sigma_t^2$ is a predefined variance schedule and the denoiser $\boldsymbol{\mu}_\theta$ is parameterized by a neural network.

Denoising in pixel space is inefficient and cannot scale up to high resolution. To address this issue, Latent Diffusion~\shortcite{Rombach_2022_CVPR} proposes to denoise in the latent space. More specifically, a VAE is first trained to compress images into latent space, and then the diffusion denoiser is trained to denoise in the latent space. The result of the denoising process is decoded back to pixel space by the VAE decoder.

\subsubsection{Architecture}
Our model builds upon the architecture of Animate Anyone, as empirically we find it achieves a better balance between performance and efficiency than other concurrent methods. The inputs are an image of the target character for appearance reference, and an image showing the skeleton in the target new pose (Fig.~\ref{fig:networkinputs}.a). The skeleton image is encoded using a Pose Guider and fused with noisy latents as inputs to the Denoising UNet. 
The appearance reference image is fed through CLIP image encoder to produce semantic features for the cross attention layers. It is also fed through a ReferenceNet to produce spatial features, replacing the self-attention in the UNet.

Our model defaults to working on a white background, and we trained a useful variation that works on other backgrounds. Animate Anyone and most other existing methods assume the background of the reference appearance image is the same as what is desired for the output. However, in many real-life videos, the background is also moving. To reduce this constraint, we use Grounded-SAM~\cite{ren2024grounded} to segment the foreground object and place it onto a white background following our default input format of appearance reference. Then we use Remove Anything~\cite{yu2023inpaint} to inpaint the background. To add the condition of background, we find it effective to extract features with the same ReferenceNet and add them to the spatial attention module. In this way, the user could use a background for the target new pose that is different from the one in the reference image. Fig.~\ref{fig:more_results} shows results with different backgrounds.

\subsection{Training Data Synthesis}
\label{sec:traindata_gen}
\begin{figure}[h]
    \begin{center}
        \includegraphics[width=1\linewidth]{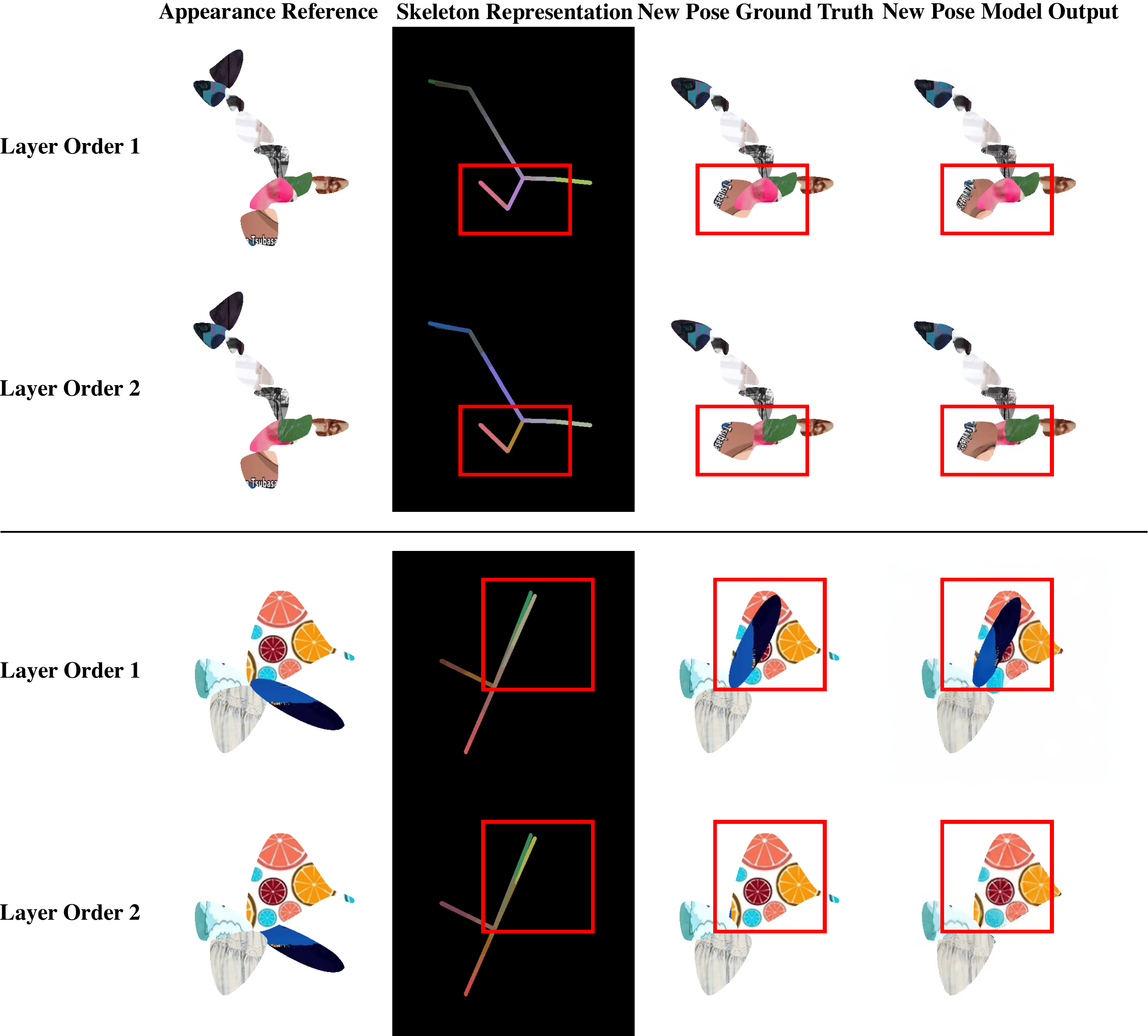}
    \end{center}
    \caption{\textbf{More Training Data Visualizations.} For each canonical appearance, we show one target pose and two layer ordering examples. The red boxes highlight how ordering affects the bone colors in skeleton representations and target appearance.}
    \vspace{-15pt}
\label{fig:more_train}
\end{figure}

What distinguishes us from other works is the data generation and model training pipeline. Acquiring enough annotated data for the ambitious goal of animating arbitrary topology is very expensive, so we develop a synthetic dataset that generalizes well to real cases during test time. Our pipeline generates a pair of appearance image and its corresponding skeleton image at a time. For each training iteration, we generate two pairs of (appearance, skeleton) image. We first generate a character at a rest pose, and the appearance image of this pair will be used as the reference image for the network. Then we randomly deform it to a new pose. For this second pair, the skeleton image of this pair will be used as the skeleton input to the network, and the appearance image would be the desired output, so it would be used as the ground truth to compute the training loss. We now detail the generation process of such paired data.
\subsubsection{Topology}
 We first generate a random tree structure to be the topology of our abstract character: we generate nodes of a random quantity within a predefined maximum, and for each node randomly assign a parent node. Then we plot the tree graph with random edge length. As the appearance reference should ideally provide as much texture information as possible, it is beneficial to avoid occlusions of body parts, and thus we define the ``rest'' pose of an arbitrary topology character as the force-directed layout. In fact, for humanoid topologies, this layout is close to the commonly-used T-shape rest pose, which supports the validity of our design choice. 
\subsubsection{Texture}
Given a tree structure, we connect each node and its parent to form an enclosed shape. As there is only one root node, we associate each shape to the child node, and the root node has no association. Suppose one axis of the shape is the line connecting the node and its parent, then we sample several points within a random aspect ratio in the direction orthogonal to the axis, and on both sides of it. With these control points, we draw Bezier curves to form a random shape. Parameters for the Bezier curves, such as the radius and the maximal number of random control points, are predefined. The back-to-front ordering of these shapes are randomly generated, and the complete character is acquired using alpha composition following that order.
We use the enclosed shape as a mask over a random image from our texture image dataset to produce the final textured blob. To let our model handle both real-life textures and more artistic or cartoon styles, our texture dataset includes at least 10000 random image samples from Cartoon Classification~\cite{cartoon_cls_dataset}, LAION Art~\cite{schuhmann2022laion} and MSCOCO~\cite{lin2014microsoft} (only raw images from these datasets are needed).

In summary, rather than validating sampled characters, our generation procedure guarantees desired properties by construction: (1) each part is a closed shape that spans the length of its corresponding bone, and its oriented bounding box has a bounded aspect ratio. (2) The topology is a tree. (3) There can be overlapping parts, but using force-directed layout as the canonical pose minimizes this in practice for the appearance reference.
\subsubsection{New Pose}
To form the second pair of data in a target new pose, we randomly transform the first rest-posed pair. We randomly select several branches of the graph, and rotate each shape in that branch by a random degree. Then the entire character is rotated by a random degree and translated by a random vector. 

\subsection{Skeleton Representation}
\label{sec:skeleton_rep}
Previous works focusing on a specific category of characters, such as humanoids or quadruple animals~\cite{xu2024animatezoo} assume the number of keypoints is fixed, and use a predefined color to draw each joint and bone. Therefore, the model may exploit the association between a specific color and the corresponding body part. For example, in skeleton images drawn by OpenPose~\shortcite{cao2017realtime,wei2016cpm,simon2017hand,8765346}, the red color always corresponds to the right shoulder of a person. However, as we are not targeting at any specific topology, we discourage the binding of color with semantic meaning. On the other hand, we hope the color provides information about the spatial transformations from the rest post to the target new pose. Therefore, we represent the position of each pixel with a linear mapping from its coordinate to a color value. Given an image of resolution $(A, B)$, and a point of coordinate $(x, y)$, our design is to color the bone at that pixel with RGB value $(c*\frac{x}{A}, c*\frac{y}{B})$, where $c$ is some scaling constant. For example, for an image of size $255\times255$ (Fig.~\ref{fig:networkinputs}), we can set $c=255$ and color a skeletal bone passing through a pixel at $(x, y)$ directly by setting the Red channel color value to $x$, and Green channel to $y$, in uint8 RGB space. 

For the Blue channel, we encode the layer ordering into it. We evenly divide the color space, and assign a value to each bone of the skeleton based on its layer index. For example, if there are five parts at most, then the back-most bone would have blue value $255/5=51$, and the foremost bone $255$. The background is set to all zeros such that its value never overlaps with a potential bone color. The final RGB skeleton representation can be seen in Fig.~\ref{fig:networkinputs}.a.

For the second pair at a new pose, instead of re-drawing the skeleton image with the updated position, we transform the bones in the rest pose to build the spatial correspondence that could be informed now by the RGB values: the current position of the bone is the desired new pose, yet its color implies where it comes from.

\section{Training and Implementation}
\begin{figure*}[h!]
\begin{center}
\includegraphics[width=1\linewidth]{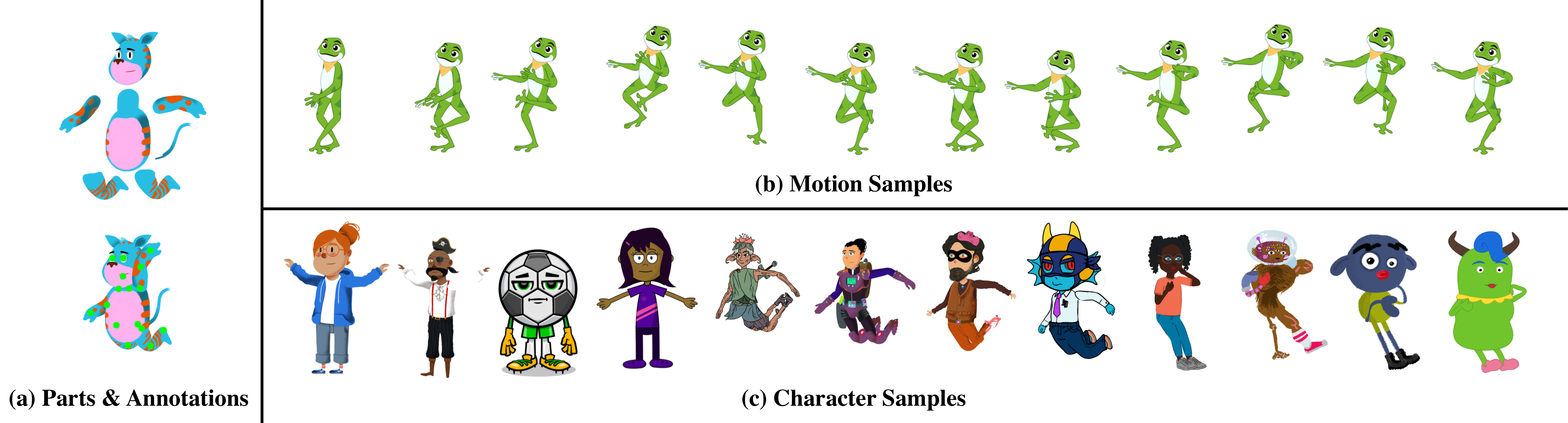}
\end{center}
    \caption{\textbf{AniDiffusion Dataset.} We establish the first 2D animation dataset with accurate keypoint annotations, part segmentations and alpha masks (See (a), where keypoints are labeled in green). We use Adobe Character Animator to create more than 120 characters (c) with approx. 100 types of motion for each (b).}
\label{fig:dataset_vis}
\end{figure*}

As there are $3\times10000=30000$ texture image candidates, and we set the maximum number of bones in a topology to be 10, the combination of texture, topology and blob shapes is far beyond $30000\times10!=3.6288e10$, while we only train for 30000 steps for our experiments. The high diversity of training data prevents model memorization. The model must learn to correctly decode the information in our skeleton representation, bind each bone to the corresponding textured shapes, and transform the shapes based on the conditioning images.

For each abstract character, we generate multiple target poses and multiple layer orderings to facilitate model understanding. As in Fig.~\ref{fig:networkinputs}.b, when there is occlusion for the same new pose, the desired appearance is different when the layer ordering changes. The model has to figure out the cause of this difference to further reduce training loss.

Our training involves two stages. In Stage 1, we train for 30000 steps on 2 A100 GPUs. The learning rate is 1e-5 and the image resolution is $512\times512$. Both ReferenceNet and Denoising UNet are initialized from Stable Diffusion v1--5~\cite{Rombach_2022_CVPR}, and only the Pose Guider, the Denoising UNet and the ReferenceNet are tuned.

After Stage 1, our model can already re-pose unseen test character zero-shot, even though real-life characters look very different from the training abstract shapes to human eyes. To improve its performance, we fine-tune it on a few frames in Stage 2. Note that some of the top-performing works focusing on humanoids is able to get rid of this step, yet it is necessary in our case due to the much more relaxed assumption of the character appearance and topology. As shown in Fig.~\ref{fig:networkinputs}.a, we require the user to provide a few frames with annotated keypoint coordinates (i.e. joints), the connections of these joints (i.e. bones), and the layer ordering of the bones. We have built a front-end for this annotation process, for which please refer to the supplemental. These are the only extra inputs we need from the user, as our pipeline will then automatically plot the skeleton image for each annotated frame, and start the fine-tuning with Stage 1 model weights. Stage 2 training procedure is the same as Stage 1 but with different hyper-parameters: we use 2000 training steps here.

\section{Evaluation}
\label{sec:results}

For better evaluation of this novel task, we establish a dataset of 2D characters with keypoint annotations, which is composed two types of contents. The first component is our AniDiffusion Dataset, which contains 135 characters with various poses, and, to the best of our knowledge, is the first cartoon dataset that provides per-frame accurate keypoint annotations and alpha mattes for the character. The data generation software is Adobe Character Animator (Ch), which comes with rich character and motion libraries. The topologies of many characters resemble humans, but the appearances are much more varied, as shown in Fig.~\ref{fig:dataset_vis}.a. There is a Motion Library that defines more than 140 motions for each character, ranging from walking to fighting (Fig.~\ref{fig:dataset_vis}.c), and the human joint definitions correspond to other widely-used packages such as OpenPose~\shortcite{cao2017realtime} (Fig.~\ref{fig:dataset_vis}.a). For other characters that do not have motions programmed, we script the software to pose each joint at multiple evenly spaced angles with respect to its parent joint, and permute over all angle combinations of available joints. We also render each primitive body part of the character in separate for each frame, to clarify the occluded regions, and enable more flexible compositions, such as adding a tail from an elephant character to a cartoon human. Finally, we provide 48 background images that could be composed with the character RGBA image.

As Character Animator focuses on animating humanoid characters, the second component of our evaluation dataset are videos containing objects of more diverse topologies and appearance from the Internet, ranging from insects and marine creatures, to machines and toys. We use a mixture of Co-tracker\cite{karaev2023cotracker} and manual annotation to label keypoints. The keypoint locations are selected such that the major motion could be described concisely by the resulting skeleton, and are verified by the authors. We use the first frame as the appearance reference image, and two frames, the middle and the ending one of the sequence, as the fine-tuning frames. The video length spans from 5 to 67 frames. As the selected clips contain non-repetitive motions that span over the entire sequence, usually these three frames have distinguished poses and the test poses are smooth interpolations of them (a small portion would be mild extrapolations). This fixed selection strategy lets us use informative fine-tuning samples without much cherry-picking.

\subsection{Qualitative Evaluation}
\begin{figure*}[t]
    \begin{center}
        \includegraphics[width=1\linewidth]{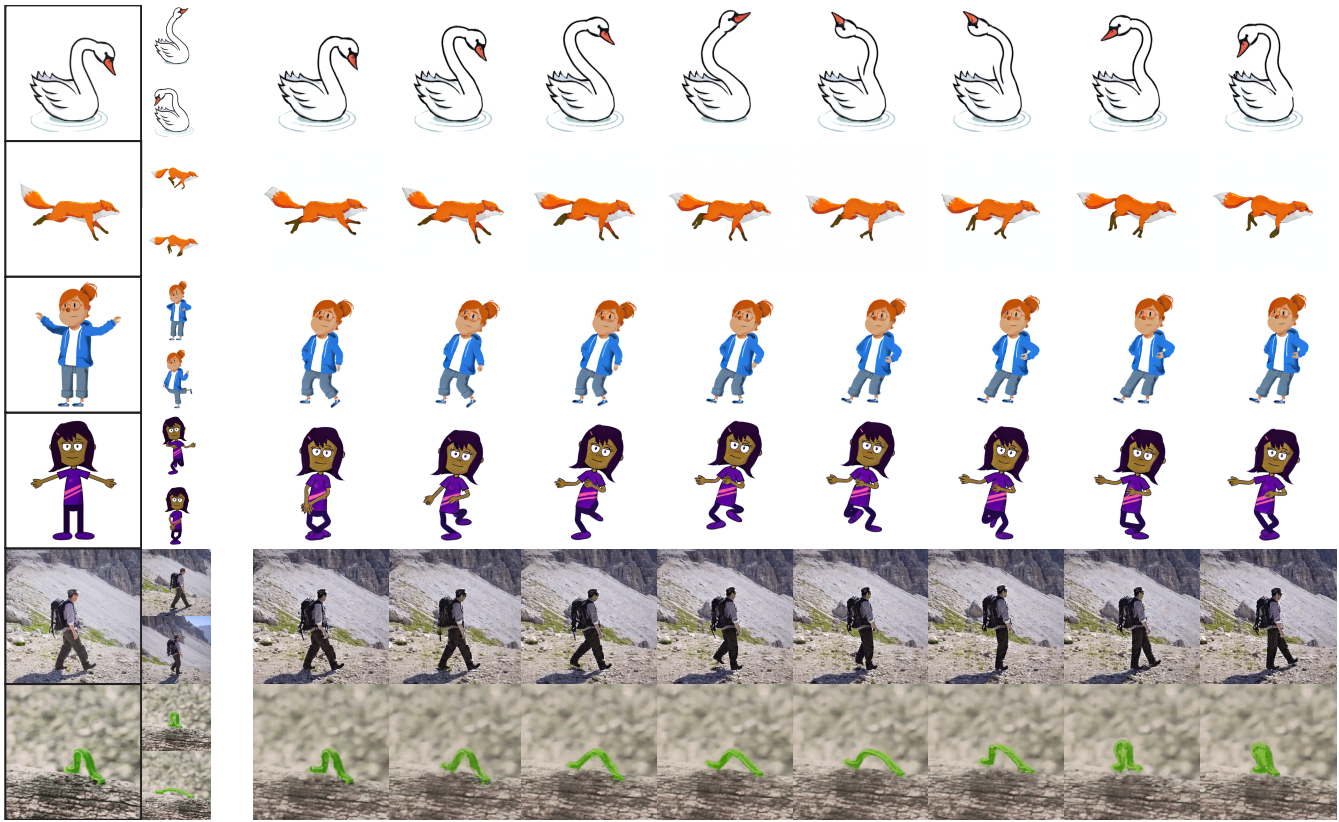}
    \end{center}
    \caption{\textbf{Result Visualizations.} In the left most column, we show the reference image, the only two fine-tuned frames (in thumbnails). On the right we show equal-spaced consecutive frames from our model outputs. After a 25-minute fine-tuning on only these three frames, our results show impressive identity preservation, motion interpolation quality, and temporal coherency. From cartoons (row 1--4; row 3--4 are results on AniDiffusion dataset), to real life clips (row 5--6), our model works on a wide range of contents and styles. Please see our supplemental materials for more examples. Image credits (row 1,2,5,6): Fisherfield Childcare, Edina Gecse, DAVIS-2017~\cite{Pont-Tuset_arXiv_2017}, Arianna1 $@$ Tenor.}
\label{fig:more_results}
\end{figure*}

We start by showing sequential results of a wide range of topologies, styles, and motions in Fig.~\ref{fig:more_results}.
\begin{figure*}
\begin{center}
    \includegraphics[width=1\linewidth]{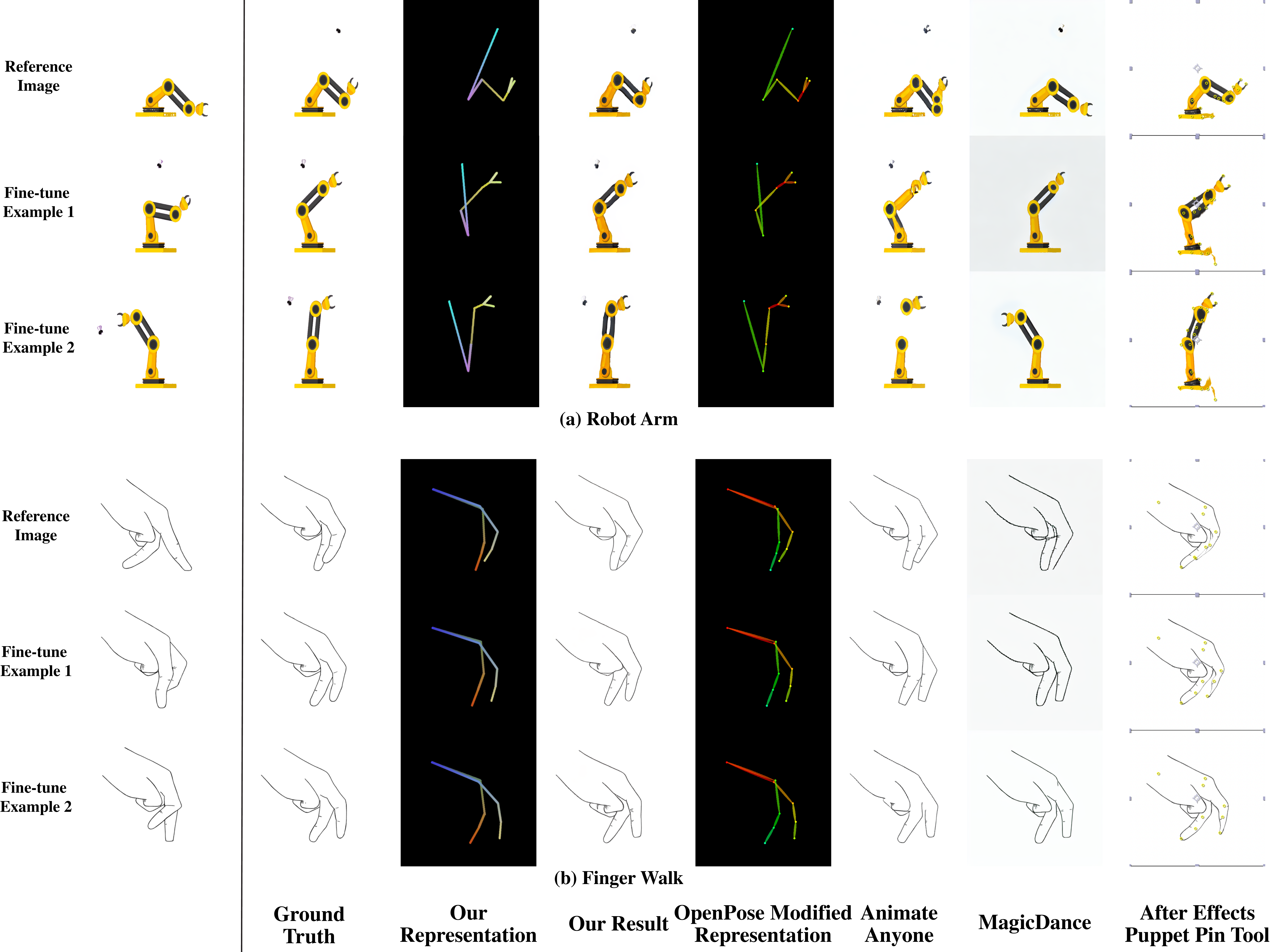}
\end{center}
    \caption{\textbf{Qualitative Comparisons.} We compare with two top-performing pose-conditioned diffusion methods, Animate Anyone and MagicDance, and one standard editing tool powered by multiple classical deformation algorithms, Puppet Pin Tool in Adobe After Effects. The bone representations are shown on the left to the matching results, and are the same for Animate Anyone and MagicDance. Image credits (top to bottom): George Girgis, NobleDame $@$ Tenor.}
\label{fig:qual_comp}
\end{figure*}

Then we compare with two state-of-the-art pose-conditioned diffusion methods, Animate Anyone~\shortcite{karras2023dreampose} and MagicDance~\cite{chang2023magicdance}. As there are no existing methods that target for arbitrary topologies, we need to modify the skeleton representation and fine-tune these methods. To reveal the fundamental ability of the re-posing components, we keep the comparison fair by omitting training of the temporal component, and use the same seeding for all. As Animate Anyone and MagicDance use OpenPose skeleton images targeted at humans, we modify the plotting function such that it could map each keypoint in a video to a fixed color for all frames. 

Fig.~\ref{fig:qual_comp}.a shows a robot arm catching a fly. The fly is detached from the arm and is not yet into the picture in the first frame, which we used as the reference image. The longest line in our pose image points from the arm base to the fly, and therefore extends to the edge of the canvas when the fly is absent. The deformations of the robot is mostly rigid. The texture mapping from the reference to the skeleton for Animate Anyone is poor, whereas MagicDance overfits to the 3 training poses and could not smoothly interpolate to generate new poses, and could not synthesize small objects like the fly consistently. 

Fig.~\ref{fig:qual_comp}.b is a sketch example of finger walking. This is a very challenging task as the two moving fingers have similar appearances, and accurate texture-skeleton binding becomes necessary. As Animate Anyone lacks this ability, it gets the order wrong for the middle row, and omits the separating line that should tell the order for the first and the third row. MagicDance is better at this example, yet its generation quality is not as high, as in the first row. 

In the rightmost column of Robot Arm, we show our best efforts to deform the reference image to the target poses using the Puppet Pin Tool in After Effects, which is powered by a collection of widely-used classical deformation methods. The yellow circles are the placed pins. It always needs more control points than neural methods, and the set of necessary points are different for different poses, which means the user needs to ``overfit'' to each pose through lots of trial and error. There are also many distortion artifacts caused by the fact that there is no easy way for classical methods to separate out multiple layers from an image input. For Finger Walk, we instead tried to use the same set of keypoints that were used for neural methods. Severe distortions occur near the finger tips. As there are no longer enough pins to fix certain parts, the displacement of one point incorrectly influences too many of its neighboring pixels.

 \subsection{Smooth Interpolation for Non-Rigid Deformations}
\begin{figure}[h!]
\begin{center}
    \includegraphics[width=1\linewidth]{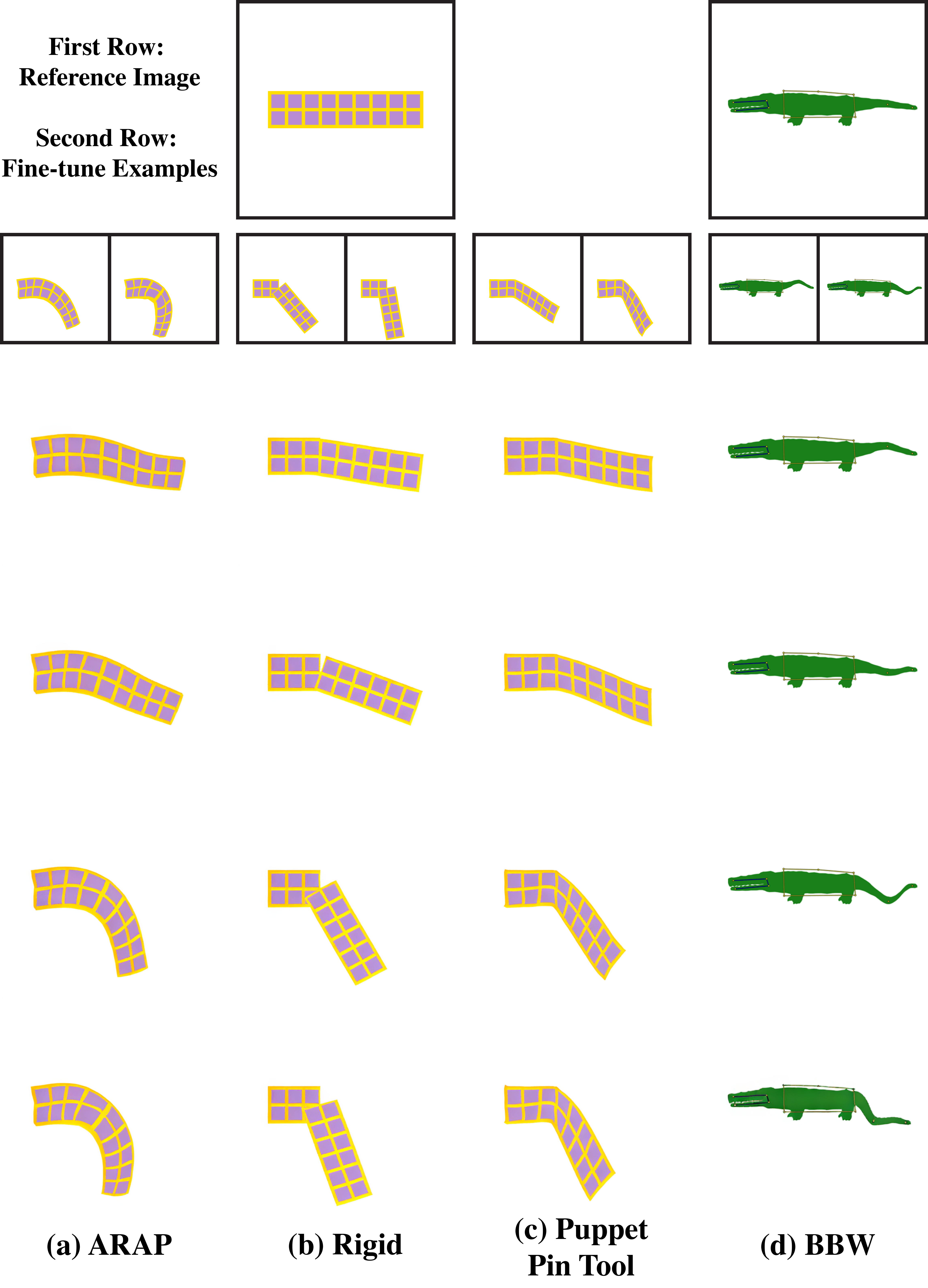}
\end{center}
    \caption{\textbf{Interpolation Results.} Our model interpolates between fine-tuned examples smoothly regardless of the underlying deformation algorithms. Figure (a)--(c) are self-created, and (d) is from the original BBW paper~\cite{jacobson2011bounded}.}
\label{fig:smooth_interp}
\end{figure}

\begin{figure}[h!]
\begin{center}
    \includegraphics[width=1\linewidth]{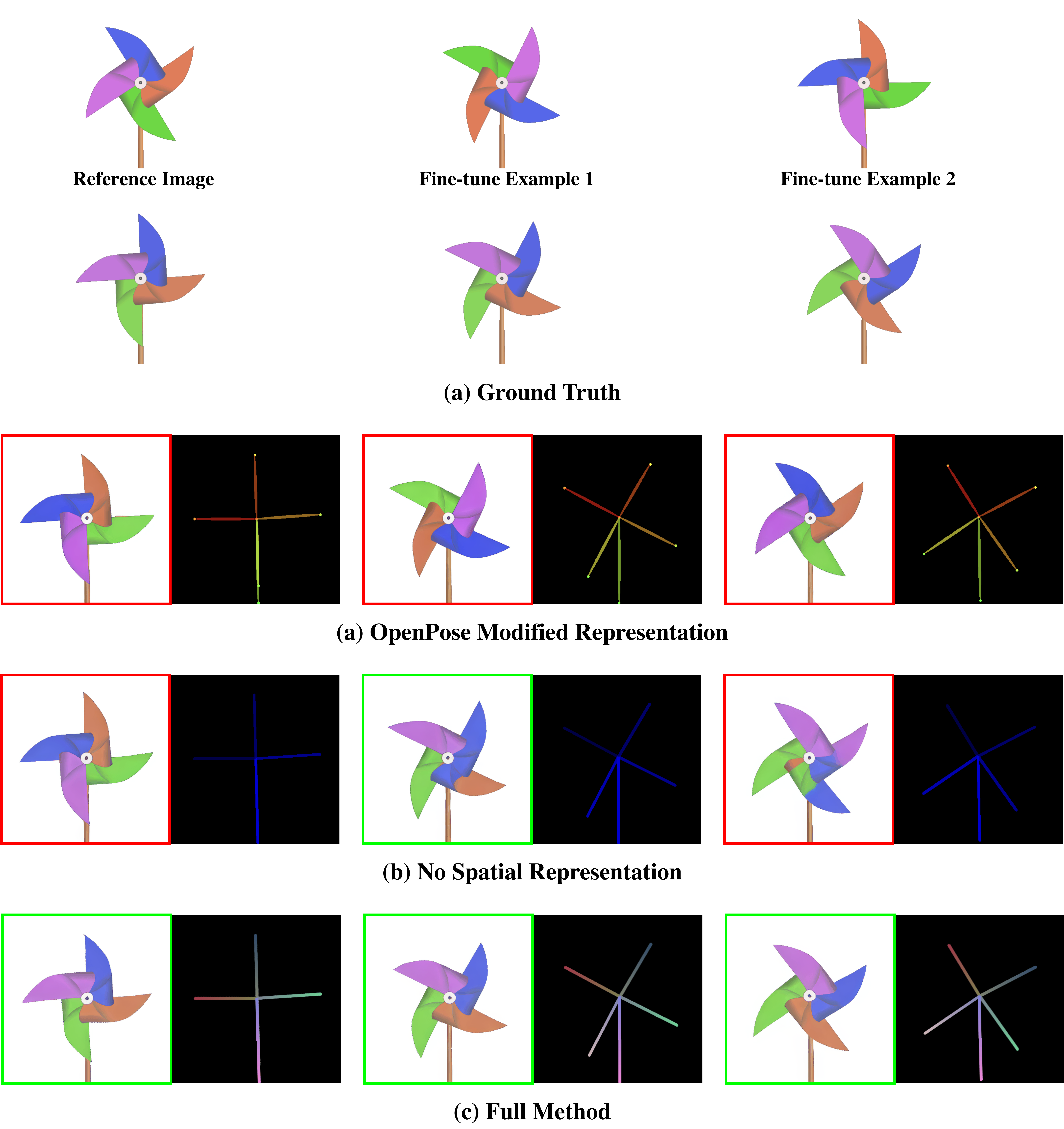}
\end{center}
    \caption{\textbf{Ablation of Skeleton Representation.} Correct results are marked by green frames, and incorrect ones by red. Our full method generates all three images correctly, the ablated version generates one correctly, and Animate Anyone memorizes and repeats the fine-tuned examples as the outputs, getting all three wrong. Image credits: Ekaterine Kantaria.}
    \vspace{-5pt}
\label{fig:ablation_pinwheel}
\end{figure}

The Stage 1 synthetic data only contains rigid transformations but not character-specific details. In the second fine-tuning stage, we let the model adapt to the specific, complex skinning effects of the target test character. One of the key observations is that diffusion models possesses the surprising power to interpolate reasonably no matter what underlying algorithm was used to create the input images. In Fig.~\ref{fig:smooth_interp} we provide an essential demonstration by bending the same grid rectangle with different classical algorithms: (a) ARAP~\cite{sorkine2007rigid}, (b) rigid rotation and (c) Puppet Pin Tool. We fix the control points, and therefore our model takes in identical skeleton images, and must reversely infer the deformation mechanism from the three fine-tune examples (one is the rest-pose appearance image, and two in other poses as shown in the first two rows). For BBW~\cite{jacobson2011bounded}, due to difficulties in Python re-implementation, we use the demo example, an alligator image, in the original Matlab code. We deformed the tail with a control point defined in the paper. Our model adapts to a reasonable deformation pattern for all methods. Although the generation might not be strictly following that particular algorithm, the visual quality is already reasonable, and we leave more thorough quantification of this ability as an interesting future investigation. 

\subsection{Quantitative Evaluation}
\begin{table}
  \centering
  \begin{tabular}{c|c|c|c|c}
    \toprule
    Method & MSE$\downarrow$&PSNR$\uparrow$&FID$\downarrow$ &LPIPS$\downarrow$\\
    \midrule
    MagicDance & 15.51&1.64&294.44&9.03\\
    Animate Anyone & 12.44&1.88&3.35&4.73 \\
    Ours & \textbf{8.61} & \textbf{2.04}&\textbf{2.87}&\textbf{3.77}\\
    \bottomrule
  \end{tabular}
  \caption{\textbf{Quantitative Comparison.} The scale of MSE, PSNR, FID and LPIPS is 1e-3, 1e+1, 1e-2, 1e-2. The best results in the comparison with other methods are highlighted.}
  \label{tab:quant_comp}
\end{table}

We run quantitative evaluation on a subset of our data with white backgrounds. As shown in Table~\ref{tab:quant_comp}, we use common metrics including Mean Squared Error (MSE), Peak Signal-to-Noise Ratio (PSNR), Fréchet Inception Distance (FID)~\cite{heusel2017gans} and Learned Perceptual Image Patch Similarity (LPIPS)~\cite{zhang2018unreasonable}. While MSE and PSNR measure the absolute difference between the predicted frames and ground truths, FID and LPIPS offer a complementary view by using trained networks to measure perceptual differences and diversity. AniDiffusion beats other methods in all.

It is noteworthy that we fine-tune Animate Anyone using the same setup as ours due to model similarity. However, for MagicDance we need to train it for 10000 steps for each of its 2 stages. The detailed comparison of computation cost is in Table~\ref{tab:compcost}, which shows that AniDiffusion achieves impressive results while staying lightweight.

\begin{table}
  \centering
  \begin{tabular}{c|c|c|c|c}
    \toprule
    Method & \thead{FT T\\(min)}&\thead{FT M\\(GB)}&\thead{Infer T\\(min)}&\thead{Infer M\\(GB)}\\
    \midrule
    MagicDance & 356.93&32.25&9.59&47.41\\
    Animate Anyone & 23.29&24.25&3.73&12.96\\
    Ours & 28.50& 24.23&2.94&12.98\\
    \bottomrule
  \end{tabular}
  \caption{\textbf{Runtime and Memory.} T refers to total time, and M refers to maximal memory. FT refers to Stage 2 fine-tuning on the target test character, and Infer refers to generating the entire video based on per-frame skeleton image inputs. For both fine-tuning and inference, MagicDance takes significantly more time and memory, while Animate Anyone and our method are roughly on the same level. This is expected because the major differences between our work and Animate Anyone are in the procedural synthesis and skeletal representation of pose. The measured fluctuations may be due to nuances in the implementation, or other factors of the server where our GPUs are hosted upon.}
  \label{tab:compcost}
\end{table}

\subsection{Ablations}
\begin{table}
  \centering
  \begin{tabular}{c|c|c|c|c}
    \toprule
    Method & MSE$\downarrow$&PSNR$\uparrow$&FID$\downarrow$ &LPIPS$\downarrow$\\
    \midrule
    w/o SE & 8.74 &2.02&3.90&3.63\\
    FT 1& 10.55&1.95&3.47&4.41\\
    FT 2 Seq&13.78 &1.86&11.06&5.99\\
    \midrule
    -80\% & 9.49 &1.98&5.28&4.20\\
    -60\% & 9.94 &1.98&3.99&4.13\\
    100\% & 8.16 &2.04&2.87&3.77\\
    +60\% & 7.55 &2.98&3.21&3.40\\
    +80\% & 7.90 &2.07&3.74&3.78\\
    \bottomrule
  \end{tabular}
  \caption{\textbf{Ablations.} In the top three rows, w/o SE refers to our skeleton representation scheme without the Red and Green channel spatial encoding, FT 1 refers to fine-tuning our complete method on only one new pose example, and FT 2 Seq refers to fine-tuning on two consecutive new poses. In the bottom rows, the first column marks the amount of training data change compared to our default setup ($100\%$), which has been reported in Table~\ref{tab:quant_comp}.}
  \label{tab:ablation}
\end{table}
For ablations, we turn off one key component at a time. We first omit the spatial information embeded in the Red and Green channel of our skeleton images. The Blue channel is still able to differentiate bones based on layer ordering, but the correspondences get poorer (see w/o SE in Table~\ref{tab:ablation} and Fig.~\ref{fig:ablation_pinwheel}). Our results are all generated by fine-tuning on three images: one reference and two new poses. If we take one pose out, the performance decreases unsurprisingly (Table~\ref{tab:ablation} FT 1). If we still use two poses, but choose sequential ones (equally spaced by 1 frame between each and from the reference image pose) that cover much less of the pose space, the performance decreases even more (Table~\ref{tab:ablation} FT 2 Seq). The diversity of the fine-tuned poses is more important than pure quantity. We also ablate on the amount of training data used. Before reaching the 30k data that we used throughout the paper, the metrics improve visibly with more training data. However, adding more synthetic data beyond 30k is no longer very helpful, even though ample synthetic data is available. These diminishing returns suggest the model has already learned most of the underlying patterns. Higher model capacity and image resolution may bring extra improvements, which we leave as future work.

\section{Discussion}
Currently our model only works in 2D. Its relaxed assumption on object appearance comes at the cost of the inability to predict contents unseen in the fine-tuning examples. For example, if the reference appearance image shows the front of a person, we cannot simulate the person turning around. Another improvement that could greatly expand our use case is to combine other forms of conditioning, as we cannot control contents that are not captured by the skeleton and the reference image (please see supplemental materials for example). Our task is challenging and our results still contain some artifacts. The method could be further optimized by incorporating helpful modules such as temporal attention~\cite{guo2023animatediff} and faster sampling~\cite{huang2023decoupled}. We believe these are interesting future directions.

\section{Conclusion}
This work presents a pioneering step towards taming the powerful diffusion models for pose-controlled diverse cartoon generation. We propose a rigging model with great generalizability to a wide range of textures and topologies, and AniDiffusion Dataset, the first cartoon dataset with accurate joints annotations to facilitate future related research. With extensive experiments, we demonstrate why adapting existing humanoid-focused methods to this task is non-trivial, and how our method can fill this blank and has the ability to synthesize high-quality re-posed cartoon characters.

\newpage
\bibliographystyle{template/eg-alpha-doi} 

\bibliography{main}

\end{document}